\documentclass[aps,prb,groupedaddress,twocolumn,showpacs]{revtex4-1}

\usepackage{graphicx}
\usepackage{bm}
\begin{document}


\title{Continuous Mott Transition in a Two-Dimensional
Hubbard Model}


\author{Yuki Yanagi}
\email[]{yanagi@rs.tus.ac.jp}
\affiliation{Department of Physics, Faculty of Science and Technology, 
 Tokyo University of Science, Noda, Chiba 278-8510, Japan}
\author{Kazuo Ueda}
\affiliation{Institute for Solid State Physics, University of Tokyo, Kashiwa, Chiba 277-8581, Japan}

\date{\today}

\begin{abstract}
 We investigate nonmagnetic metal-insulator transition in the
 1/5-depleted square lattice Hubbard model at half-filling within the 8-site cellular
 dynamical mean field theory.  
 We find that a metal-insulator transition 
 without any signatures of the first
 order transition, a continuous Mott transition, takes place 
 in a certain range of parameters. The nature of the continuous Mott
 transition  is nothing but a Lifshitz transition driven by the on-site
 Coulomb interaction.  The renormalized matrix elements of hoppings and
 the spin-spin correlation functions
 reveal that physics of this transition is  strong enhancement of the
 dimerization due to the non-local effects of electron-electron interaction. 

\end{abstract}




\maketitle

\section{Introduction}
 Metal-insulator transition (MIT) is one of the most intriguing
 phenomena in condensed matter physics\cite{Imada}. 
Particularly, the MIT due to the strong Coulomb interaction, 
known as the Mott transition, has 
 attracted much attention and a considerable number of theoretical and
 experimental studies have been devoted to elucidate generic
 features of the Mott transition. 
  According to these studies, 
 it has been clarified that properties of the Mott transition strongly  
 depend on the lattice structure, 
 such as connectivity and 
 geometrical frustration.    
 The Mott transition on 
   non-Bravais lattices is especially interesting\cite{Meng,Sorella,Kancharla,Ruger,Garg,Paris,Kancharla_2}. 
 In  non-Bravais lattices, 
 the crystallographic unit cell contains at
 least two distinct sublattices and 
the resulting multi-band electronic structures 
 bring various possibilities into the Mott transition.  

   One typical example is the honeycomb lattice Hubbard model 
 in which the Dirac cones are located on the Fermi level at half-filling.  
  Possibility of the Mott transition without any signatures of the
 first order transition, the continuous Mott transition, 
 and a spin liquid ground state without geometrical frustration 
  have been discussed in the literature\cite{Meng,Sorella} 
 although the recent state-of-the-art quantum Monte
 Carlo (QMC) simulation is negative for 
 the nonmagnetic insulating (NI) ground state\cite{Sorella}. 
 Another example is the bilayer Hubbard model in which the two square
 lattice Hubbard layers are coupled with each other through an
 inter-layer hopping $t_\perp$. 
 The model has been studied by 
 the cellular  dynamical mean field theory (CDMFT) with
 $\sqrt{2}\times\sqrt{2}\times 2$-site cluster\cite{Kancharla} and 
 it has been shown that for larger values of $t_\perp$, 
 the antiferromagnetic insulating (AFI) phase 
 is suppressed and a nonmagnetic metallic state appears between 
the AFI and NI phases despite the perfect nesting of the Fermi surfaces.  
  It is natural to expect that this result indicates 
  the presence of a continuous Mott transition.   
  According to the recent variational Monte Carlo (VMC)
  study\cite{Ruger}, however, the AFI phase extends for the larger
 inter-layer hopping regime and the continuous Mott transition 
  is masked by the AFI phase. 
  Possibility of the nonmagnetic metallic ground state without geometrical
 frustration and the continuous Mott transition have been discussed also in 
 the ionic Hubbard model where the staggered potential $\pm \Delta$ is
 applied on the A or B sublattices. 
 The studies by the single-site dynamical mean field theory
 (DMFT)\cite{Garg} and by the QMC
  simulation\cite{Paris} 
 give positive results, while the CDMFT study\cite{Kancharla_2}
  a negative one. 
 In any case, it is still controversial 
 whether the continuous Mott transition and the
 metallic ground state without geometrical frustration 
 is possible or not.  

 \begin{figure}[t]
  \begin{center}
   \includegraphics[width=4cm]{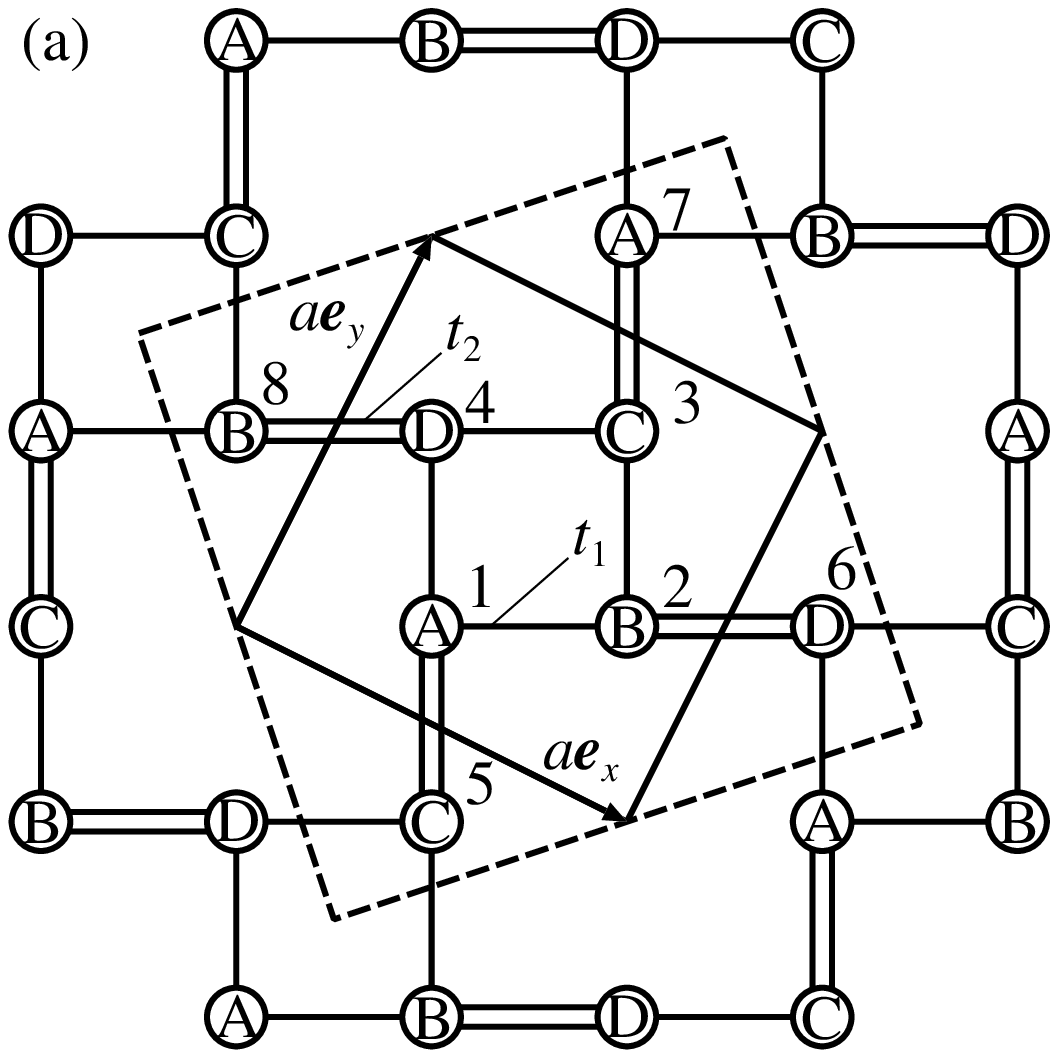} 
   \includegraphics[width=4cm]{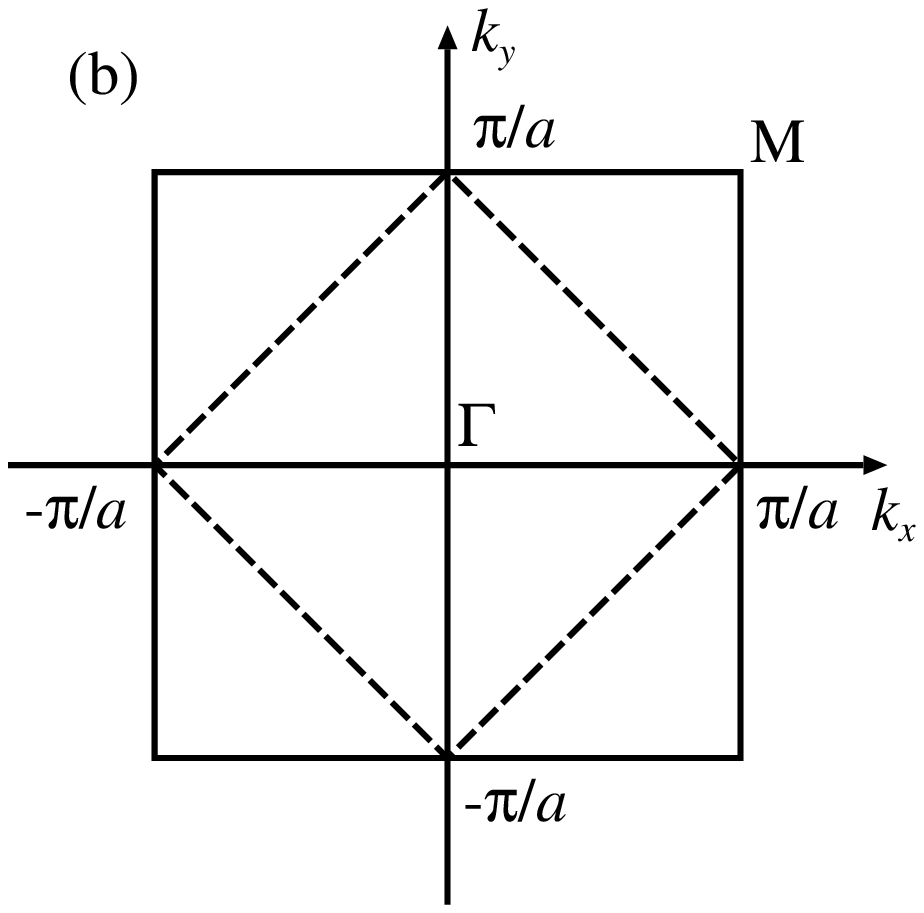} 
   \includegraphics[width=8cm]{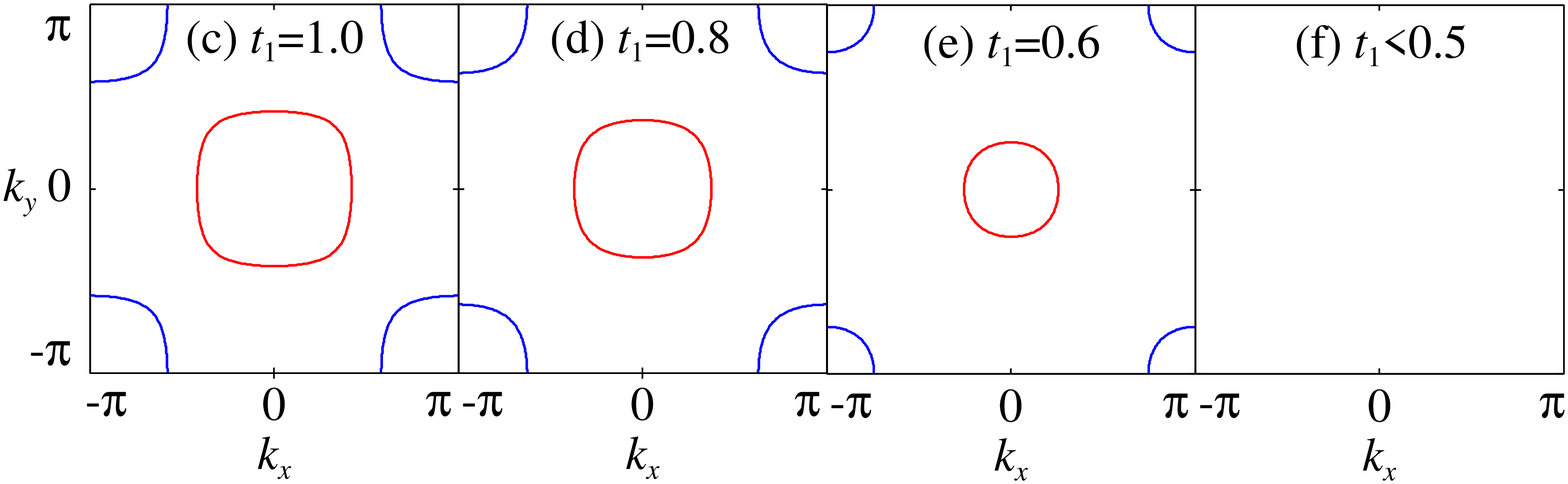} 
   \caption{(Color online) (a) The schematic of the 1/5-depleted square lattice. The
   single and double solid lines represent the intra-plaquette and
   intra-dimer hoppings $t_1$ and $t_2$, respectively. The solid and
   dashed squares represent the crystallographic unit cell 
   and the 8-site supercell, respectively. 
   (b) The first Brillouin zone (solid square) and the reduced Brillouin
   zone (dashed square) corresponding to the 8-site supercell. 
 (c)-(f) Fermi surfaces for
   several values of $t_1$.\label{fig_model}}
  \end{center}
 \end{figure}

 In the present paper, we show that 
 the Hubbard model on the 1/5-depleted square
 lattice\cite{Ueda,Troyer,Troyer_2,Yamashita,Yanagi} 
 at half-filling exhibits 
 the continuous Mott transition 
 in the paramagnetic phase, where the model in the strong coupling limit
 is considered to be a minimal model for CaV$_4$O$_9$\cite{Ueda} which is a typical
 example of spin-gapped systems.  
 The 1/5-depleted  square lattice is a kind of non-Bravias lattice in which there
 are four distinct sublattices ($A$-$D$) in the unit cell  [see
 Fig. \ref{fig_model} (a)].
 We study the MIT in the paramagnetic phase of this model 
 with use of the CDMFT\cite{Kotliar,Maier} combined 
 with the continuous-time auxiliary field quantum Monte
 Carlo method (CT-AUX)\cite{Gull_1,Gull_2}.    
   It is shown that in this model, the nonmagnetic continuous
 Mott transition is driven by the promotion of dimerization due to
 the non-local electron correlation effects.

\section{1/5-depleted square lattice Hubbard model and CDMFT}
   We consider the Hubbard model on the
   1/5-depleted square lattice,  
\begin{eqnarray}
   H=\sum_{i,j,\alpha,\beta}t_{i,\alpha,j,\beta}
   c^{\dagger}_{i\alpha\sigma}c_{j\beta\sigma}
   +U\sum_{i\alpha} n_{i\alpha\uparrow}n_{i\alpha\downarrow}, 
\end{eqnarray}
where $c^{(\dagger)}_{i\alpha\sigma}$ annihilates (creates) an electron with spin
 $\sigma$ on the sublattice $\alpha$ at the unit cell $i$, 
  $n_{i\alpha\sigma}=c^{\dagger}_{i\alpha\sigma}c_{i\alpha\sigma}$
  , $t_{i,\alpha,j,\beta}$ and $U$ represent 
   the nearest neighbor hopping integrals and
   the on-site Coulomb interaction, respectively. 
    As schematically shown in Fig. \ref{fig_model} (a), 
   $t_{i,\alpha,j,\beta}=t_1$ ($t_2$) on plaquette- (dimer-) bonds, 
where  we refer to $t_1$ ($t_2$) as the intra-plaquette (-dimer)
 hopping, here and hereafter. 
   We treat $t_1$ and $t_2$  as independent parameters  
 because they are not geometrically equivalent in the 1/5-depleted
 square lattice.  
In the present study, we concentrate on the case of $t_1<t_2$ 
and the energy is measured in units of $t_2$ hereafter. 
  According to our previous studies\cite{Yamashita,Yanagi},  
single electron properties of the model 
 are significantly different depending 
 on values of $t_1$ and $t_2$. 
 For $t_1 < t_2$, 
there are four energy bands composed of 
  two valence and two conduction bands 
and the bottom (top) of the conduction (valence)
 bands is located on the energy $-2t_1+t_2$ ($2t_1-t_2$) at the
 $\Gamma$- ($M$-) point and the total band width $W=4t_1+2t_2\le 6t_2$,
 where we note that the first Brillouin zone is usual square shape as
 shown in Fig. \ref{fig_model} (b). 
 Therefore, for $0.5t_2 \le t_1 < t_2$, the noninteracting ground
 state is (semi-)metallic and  
 there is a electron (hole) pocket around the $\Gamma$- ($M$-) point
 as shown in Figs. \ref{fig_model} (c)-(e). 
 With decreasing $t_1$, the
 Fermi surfaces shrink and finally 
 vanish at $t_1=0.5t_2$ as shown in
 Figs. \ref{fig_model} (c)-(f). 
Thus, the noninteracting ground state is insulating for $t_1<0.5t_2$. 
 
To study effects of electron correlation
in the 1/5-depleted square lattice
Hubbard model at half-filling, we use the CDMFT\cite{Kotliar,Maier}, 
one of the cluster extensions of the DMFT. 
In the CDMFT, the original lattice problem 
is mapped onto an effective
cluster problem with open boundary conditions 
embedded in a self-consistent electronic bath. 
  In our previous conference presentation\cite{Yanagi}, 
we have reported preliminary results by using the CDMFT with a
4-site cluster including 
2-dimers. The cluster has much lower symmetry than the original problem
and cluster size dependence has not been investigated. 
  In addition, for $t_1\sim t_2$, 
 the results of the Heisenberg model\cite{Ueda,Troyer,Troyer_2} indicate competition 
between the  dimer-singlet and the
  plaquette-singlet states\cite{Ueda,Troyer,Troyer_2}. 
Therefore, the CDMFT study with a larger cluster including both plaquettes
and dimers is highly desired. In the present study, 
we use the 8-site cluster depicted in
Fig. \ref{fig_model} (a).
 In the CDMFT, the bath Green's function 
 $\hat{\mathcal{G}}(i\varepsilon_n)$, the cluster Green's function 
  $\hat{G}(i\varepsilon_n)$ and 
  the cluster self-energy $\hat{\Sigma}(i\varepsilon_n)$ are
 self-consistently determined through 
the following procedure, 
 where $\hat{\mathcal{G}}(i\varepsilon_n)$, 
 $\hat{G}(i\varepsilon_n)$ and $\hat{\Sigma}(i\varepsilon_n)$
 have 8$\times$8 matrix forms and $\varepsilon_n=(2n+1)\pi T$ 
 is the fermionic Matsubara frequency. 
 Initially, one guess a  bath Green's function 
 $\hat{\mathcal{G}}(i\varepsilon_n)$. 
 Next, we solve the the effective cluster
 problem characterized by $\hat{\mathcal{G}}(i\varepsilon_n)$ with use
 of the CT-AUX\cite{Gull_1,Gull_2} 
  to obtain  $\hat{G}(i\varepsilon_n)$ and 
   $\hat{\Sigma}(i\varepsilon_n)$. 
 Then, a new bath Green's function can be computed via the CDMFT
 self-consistency condition, 
\begin{eqnarray}
&&\hat{G}(i\varepsilon_n)=\sum_{\tilde{\bm{k}}\in \mathrm{RBZ}}
 \left\{\left(i\varepsilon_n+\mu\right)\hat{1}-\hat{t}(\tilde{\bm{k}})-\hat{\Sigma}(i\varepsilon_n)\right\}^{-1}   \\
&&\left\{\hat{\mathcal{G}}(i\varepsilon_n)\right\}^{-1}=\left\{\hat{G}(i\varepsilon_n)\right\}^{-1}+\hat{\Sigma}(i\varepsilon_n)
\end{eqnarray}
where $\tilde{\bm{k}}$-integration is performed over the reduced
Brillouin zone (RBZ) shown
in Fig. \ref{fig_model} (b) and $\hat{t}(\tilde{\bm{k}})$  is the 
Fourier-transformed hopping matrix for the 8-site cluster.
This iterative procedure is performed until the converged results are
obtained. 
It is noted that the chemical
potential is $\mu=0$ at the half-filled case studied 
 in this paper because of
the bipartite nature of the 1/5-depleted square 
lattice\cite{Yamashita,Yanagi}. 
 In the present study,  
 to  investigate the  correlation effects 
 on the phase without any symmetry breaking,
  we assume a paramagnetic solution in 
 numerical calculations and set $T=0.1t_2$. 
However, we have confirmed that
 the results as will be shown below 
 do not change qualitatively down to $T=0.05 t_2$.

\section{Continuous MI transition}
 \begin{figure} [t]
  \begin{center}
   \includegraphics[width=6 cm]{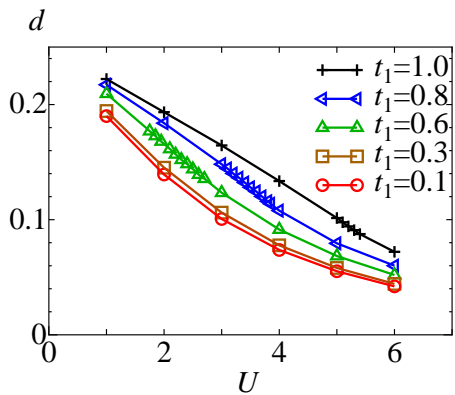} 
   \caption{(Color online) $U$-dependence of the double occupancy $d$
   for several values of $t_1$.\label{fig_docc}}
     \end{center}
 \end{figure}
 \begin{figure} [b]
  \begin{center}
   \includegraphics[height=4.0cm]{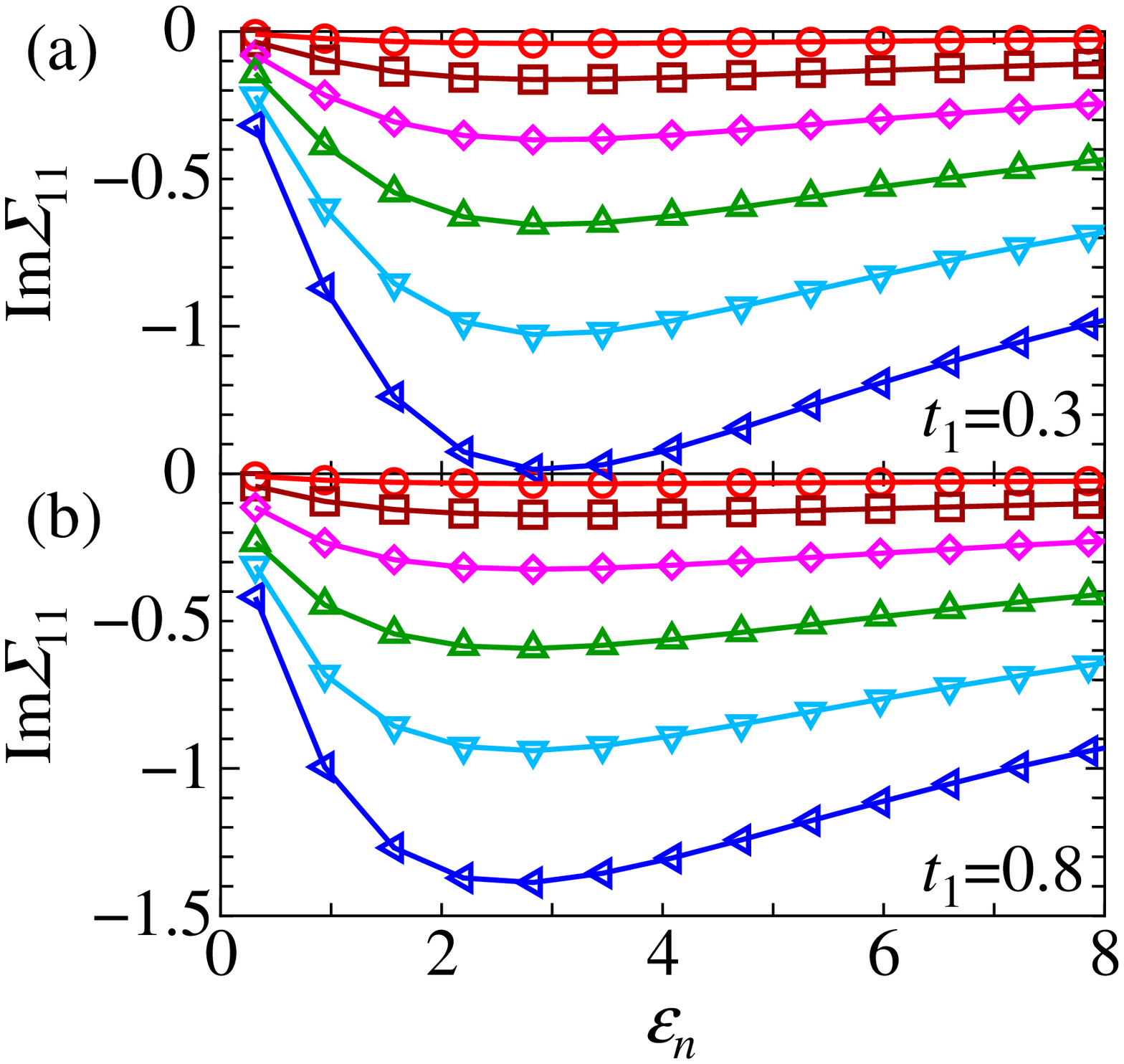} 
   \includegraphics[height=4.0cm]{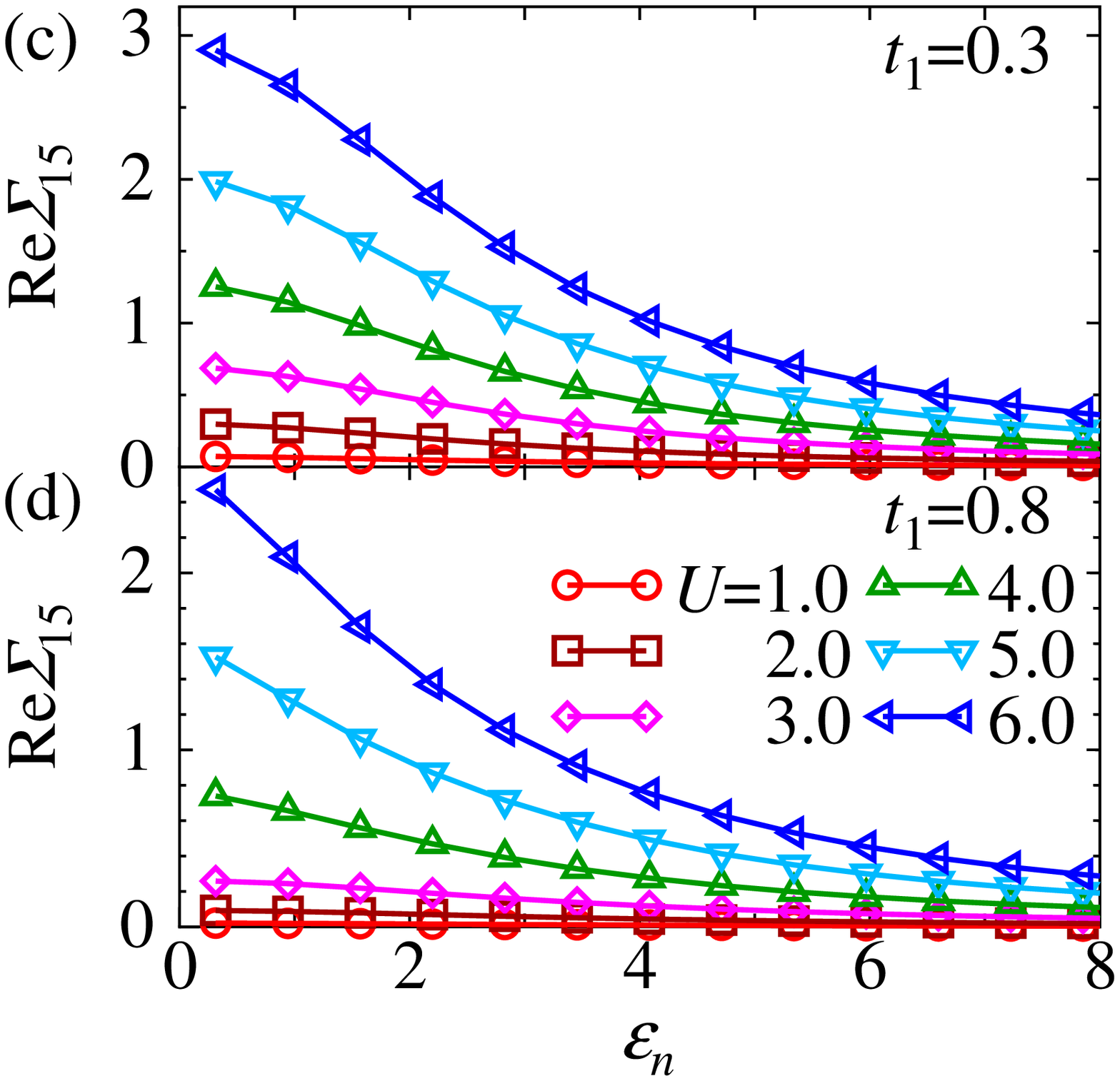} 
   \caption{(Color online) (a) and (b)  
   $\varepsilon_n$-dependence of the imaginary part of the local
   self-energy $\mathrm{Im}\Sigma_{11}(i\varepsilon_n)$, 
   (c) and (d) those of the real part of the intra-dimer self-energy 
   $\mathrm{Re}\Sigma_{15}(i\varepsilon_n)$ for $t_1=0.3t_2$ and $0.8t_2$. 
   \label{fig_docc_2}}
  \end{center}
 \end{figure}
 First, in Fig. \ref{fig_docc}, we show results on the double occupancy 
$d=\langle n_{i\alpha\uparrow}n_{i\alpha\downarrow} \rangle$ as a function of $U$
for $0.1t_2 \le t_1 \le  t_2$. 
 It is observed that $d$ decreases monotonically with increasing $U$ and 
 there is no discontinuous jump up to $U=6t_2$. This
 indicates absence of the first order Mott transition 
as far as $t_1 \le t_2$ in contrast to
 the case of the ordinary Hubbard model on the square 
lattice\cite{Park}. 
 It is noteworthy that $d$ for $t_1 >0.5t_2$ exhibits qualitatively
different behavior from that for $t_1 <0.5t_2$. 
For $t_1 <0.5t_2$, $U$-dependence of $d$ is always concave upward, 
while for $t_1 >0.5t_2$, it changes from concave downward to upward at a 
certain value of $U$, signaling the MIT as will be mentioned later.

 To further investigate the correlation effects, 
  $\varepsilon_n$-dependence of 
 the local self-energy 
$\mathrm{Im} \Sigma_{11}(i\varepsilon_n)$ and the intra-dimer
  self-energy $\mathrm{Re} \Sigma_{15}(i\varepsilon_n)$
for several values of $U$ at
$t_1=0.3t_2$ and $t_1=0.8t_2$
are shown in Fig. \ref{fig_docc_2}.  
 We note that for $t_1=0.3t_2$, 
the noninteracting ground state is a band insulator, 
 while for $t_1=0.8t_2$, that is a metal as mentioned before.
  For both cases, 
 the local components $\mathrm{Im}\Sigma_{11}(i\varepsilon_n)$ and the
 intra-dimer components $\mathrm{Re}\Sigma_{15}(i\varepsilon_n)$
 are enhanced with
 increasing $U$. 
 Especially, 
 the enhancement of $\mathrm{Re}\Sigma_{15}(i\varepsilon_n)$ in the low energy
 region is significant, 
which is the key to understand the MIT in the present model
 as will be shown later. 
 It has also been observed that 
 the local components 
 $\mathrm{Im}\Sigma_{11}(i\varepsilon_n)$ have no singularities and are
 proportional to $\varepsilon_n$ in the low energy region. 
 The analytic behavior of $\hat{\Sigma}(i\varepsilon_n)$ allows us to 
  expand  $\hat{\Sigma}(i\varepsilon_n)$ with respect to
 $\varepsilon_n$ for  $\varepsilon_n \sim 0$, 
 i.e., the quasiparticle picture 
 can be used to describe the electronic 
 states up to relatively large
 value of $U= 6t_2$. Then, the quasiparticle 
 cluster Green's function $\hat{\tilde{G}}(\omega)$ 
 is written as follows, 

 \begin{figure} [t]
  \begin{center}
   \includegraphics[width=6.0cm]{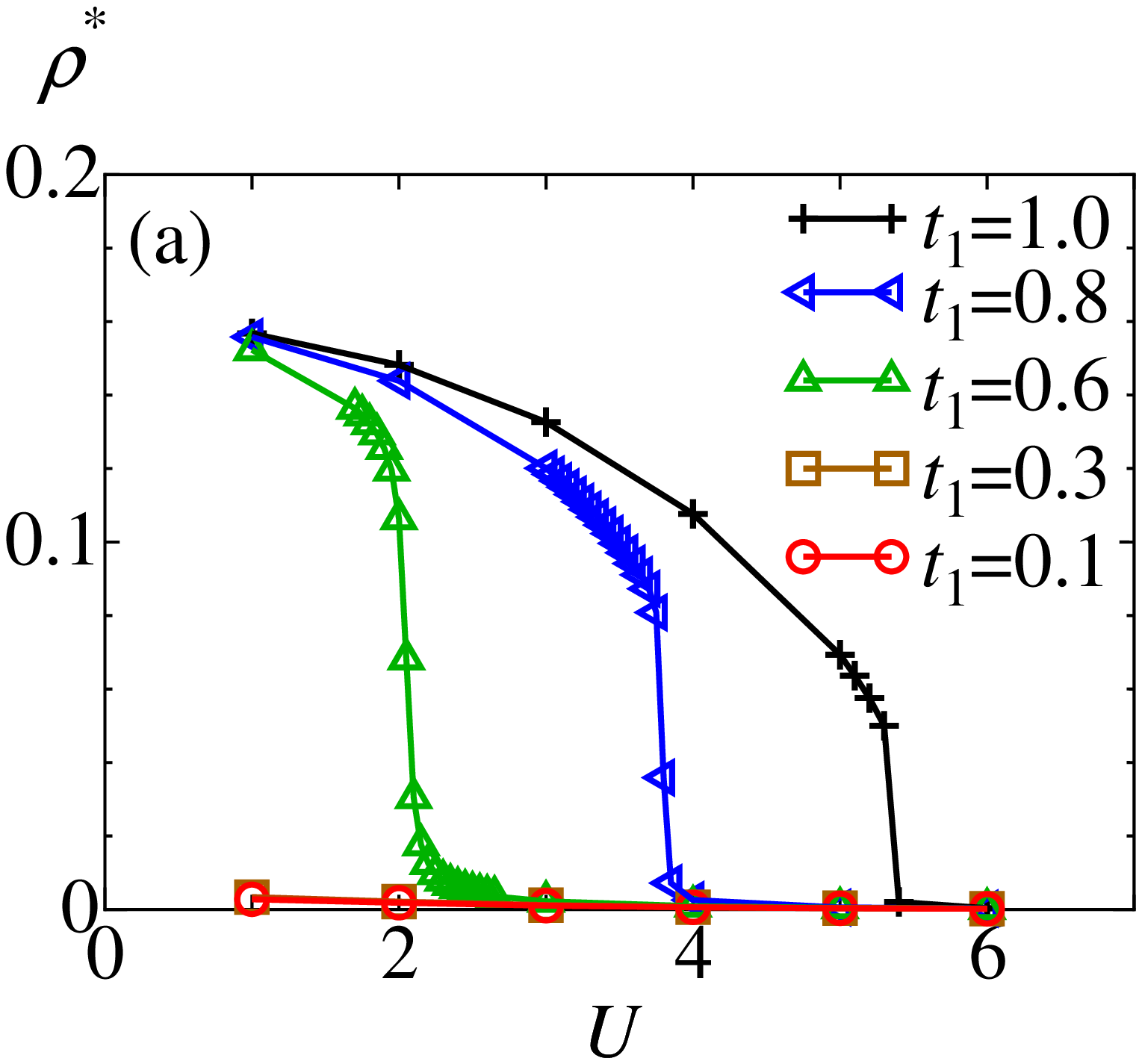} 
   \includegraphics[width=8.0cm]{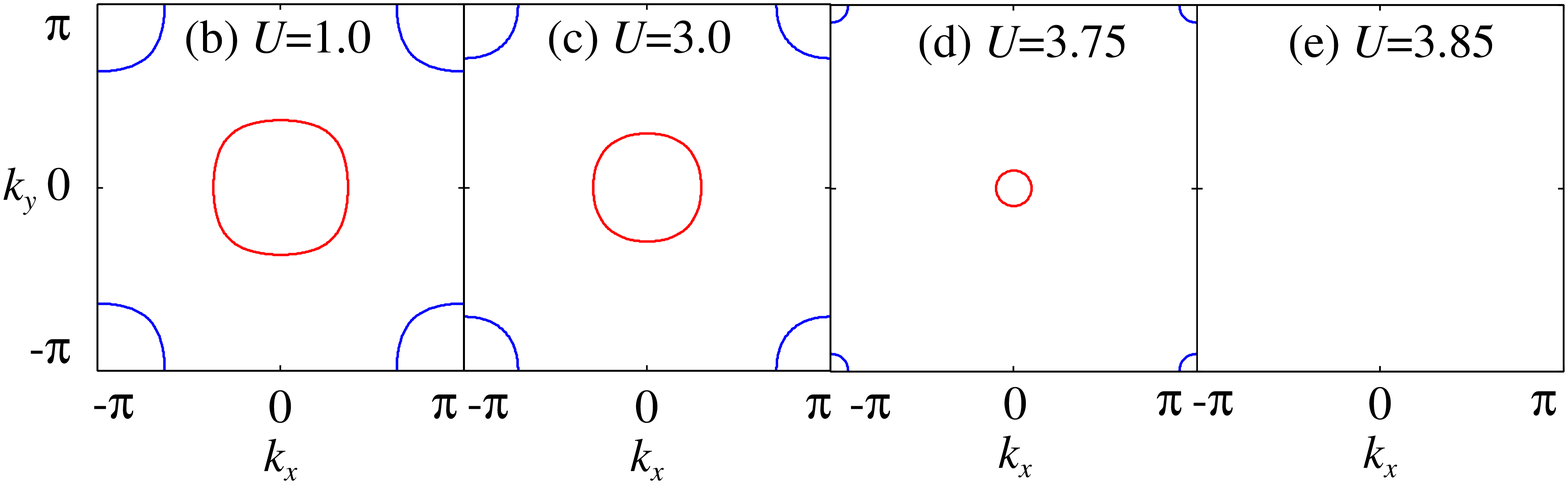} 
   \caption{(Color online) (a) $U$-dependence of the quasiparticle DOS
   $\rho^*$. It is noted that the small residual DOS is due to
   the Lorenzian broadening in the numerical calculation. 
(b)-(e) Fermi surfaces for several values of $U$ at $t_1=0.8t_2$.
 \label{fig_rho}}
  \end{center}
 \end{figure}

\begin{eqnarray}
\hat{G}(\omega)&\sim &
\hat{\tilde{G}}(\omega)=\sum_{\tilde{\bm{k}}\in \mathrm{RBZ}}
\hat{Z}^{\frac{1}{2}}\left[\left(\omega+\mu \right)\hat{1}-\hat{t}^*
		      (\tilde{\bm{k}})\right]^{-1} \hat{Z}^{\frac{1}{2}},\quad \\
\hat{t}^*(\tilde{\bm{k}})&=&
 \hat{Z}^{\frac{1}{2}}\left[\hat{t}(\tilde{\bm{k}})+\frac{\hat{\Sigma}(i\pi
		       T)+\hat{\Sigma}^{\dagger}(i\pi
		       T)}{2}\right]\hat{Z}^{\frac{1}{2}}, \\
 \hat{Z}&=&\left[\hat{1}-\frac{\hat{\Sigma}(i\pi
		       T)-\hat{\Sigma}^{\dagger}(i\pi
		       T)}{2i\pi T}\right]^{-1}, 
\end{eqnarray}
 where $\hat{t}^*(\tilde{\bm{k}})$ and $\hat{Z}$ are the  
 Fourier-transformed renormalized hopping
 matrix in the cluster and the renormalization factor, respectively.  
 The $U$-dependence of the quasiparticle density 
 of states (DOS) at the Fermi level
 $\rho^*=\frac{1}{\pi}\mathrm{Im}\hat{\tilde{G}}_{11}(0)$ 
 is  shown in Fig. \ref{fig_rho}. 
 On one hand, for $t_1 \le 0.5t_2$, $\rho^*$ is always $0$
  and the insulating ground state is realized  
  irrespective of the values of $U$. 
 On the other hand, for $0.6t_2 \le t_1 \le t_2$,
 $\rho^*$ monotonically decreases with increasing $U$ 
and vanishes at a certain critical value $U=U_c$. 
 This clearly shows that the MIT takes place at $U=U_c$ 
 without any signatures of the first order Mott transition, such as, 
discontinuous jumps in $U$-$d$ curves and singularities in
 $\hat{\Sigma}(i\varepsilon_n\sim 0)$ as mentioned before.  
In other words, the Mott transition in the present model is continuous. 

 \begin{figure} [t]
  \begin{center}
   \includegraphics[width=8.0cm]{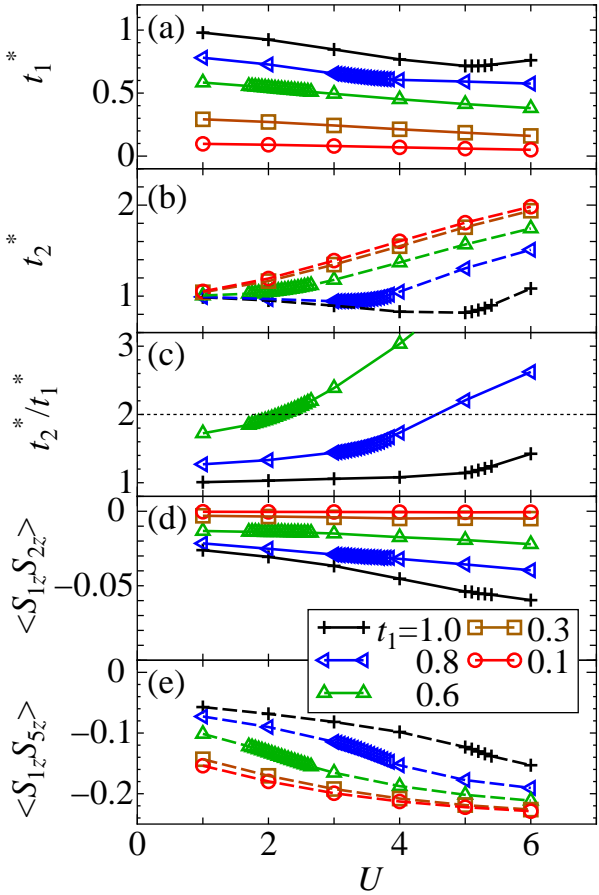} 
   \caption{(Color online) 
   $U$-dependence of (a) the effective cluster
   hopping $t_1^*$, (b) $t_2^*$,  (d) the relative value $t_2^*/t_1^*$, 
   (d) the intra-plaquette spin-spin correlation 
   $\langle S_{1z}S_{2z} \rangle$ and 
   (e) intra-dimer spin-spin correlation $\langle S_{1z}S_{5z} \rangle$
   for  several values of $t_1$. 
 \label{fig_tcl}}
  \end{center}
 \end{figure}

 To gain more insight about the MIT, 
 we have calculated the Fermi surfaces of the interacting system. 
 The Fermi surface 
in the interacting system is defined as lines of 
 the $\bm{k}$-points where the following equation is satisfied: 
\begin{eqnarray}
\mathrm{det}\left[\hat{H}_0(\bm{k})+\frac{\hat{\Sigma}^{\mathrm{L}}(\bm{k},i\pi
		       T)+\hat{\Sigma}^{\mathrm{L}\dagger}(\bm{k},i\pi
		       T)}{2}\right]=0,
\end{eqnarray}
where the Fourier representation of the kinetic part of the Hamiltonian $\hat{H}_0(\bm{k})$  and the
  $\bm{k}$-dependent lattice self-energy $\hat{\Sigma}^{\mathrm{L}}(\bm{k},i\varepsilon_n)$ have
 4$\times$4 matrix forms
whose elements 
are
labeled by the four distinct sublattices 
$\alpha,\beta$($=A$-$D$) contained in the
 crystallographic unit cell.
 It should be noted that momentum $\bm{k}$ is defined in the first
 Brillouin zone unlike $\tilde{\bm{k}}$ in eqs. (2), (4) and (5).   
Since in the CDMFT, the translational symmetry is violated, 
 to obtain the $\bm{k}$-dependent lattice self-energy 
$\hat{\Sigma}^{\mathrm{L}}(\bm{k},i\varepsilon_n)$,
 we perform the periodization,
\begin{eqnarray}
 {\Sigma}_{\alpha\beta}^{\mathrm{L}}(\bm{k},i\varepsilon_n)=
\frac{1}{2}\sum_{m,m'}\hspace{-1mm}{}^{'} 
 \Sigma_{mm'}(i\varepsilon_n)e^{-i\bm{k} \cdot (\bm{r}_m-\bm{r}_m')},
\end{eqnarray} 
where $\bm{r}_1=\bm{r}_2=\bm{r}_3=\bm{r}_4=(0,0)$, 
$\bm{r}_5=(0,-a)$, $\bm{r}_6=(a,0)$, $\bm{r}_7=(0,a)$ and
$\bm{r}_8=(-a,0)$ and the prime on the sum restricts 
the summation over $m$ and $m'$ to the sites belonging to
the sublattices specified by $\alpha$ and $\beta$, respectively, 
e.g. $m=$1 and 7 for $\alpha=A$ [see Fig. \ref{fig_model} (a)]. 

 Change of the Fermi surface 
 as a function of $U$ is shown in
 Figs. \ref{fig_rho} (b)-(e). 
 It is observed that both the electron and the hole pockets shrink
 with increasing $U$ and finally vanish at $U=U_c$ 
 in a similar way as the 
 Fermi surface evolution with decreasing $t_1$ in the noninteracting
 system which is shown in Figs. $\ref{fig_model}$ (c)-(f).  
 This suggests that the MIT for $0.5t_2 \le t_1 < t_2$ is nothing but
 a Lifshitz transition\cite{Lifshitz}. 
 It is worth mentioning that 
 a conventional Lifshitz transition 
 is topological change of Fermi surfaces 
 driven by variations of one particle parameters\cite{Lifshitz}, 
 such as a chemical potential or a transfer integral, 
 while the one we find in the  present model is driven 
 by the electron correlation effects without any
 symmetry breaking.

  Here, we show the renormalized cluster hopping 
 $\hat{t}^*=\sum_{\tilde{\bm{k}}\in \mathrm{RBZ}}\hat{t}^*(\tilde{\bm{k}})$  
 in Figs. \ref{fig_tcl} (a)-(c). 
 Since in the noninteracting case, 
 the ground state properties are determined by the values of $t_1$ and
 $t_2$ and the MIT takes place at $t_1=0.5t_2$ as mentioned before, 
  the renormalized cluster hopping 
  $\hat{t}^*$  can give us an intuitive physical picture about the MIT. 
 Figs. \ref{fig_tcl} (a) and (b) shows the $U$-dependence of
the renormalized intra-plaquette and intra-dimer hoppings 
$t_1^*\equiv t^*_{12}$ and $t_2^*\equiv t^*_{15}$, respectively.  
 We find that with
 increasing $U$, the effective 
 intra-plaquette hopping $t_1^*$ decreases due to the effects of the
 renormalization factor $\hat{Z}$ whose eigen values are smaller than
 $1$.  
 In contrast, the effective intra-dimer
 hopping $t_2^*$ decreases more slowly than $t_1^*$ 
  up to a slightly smaller value than $U_c$ 
  and starts to increase with further
 increasing $U$. 
Therefore, the resulting $t_2^*/t_1^*$
  monotonically increases with increasing $U$ as shown in
   Fig. \ref{fig_tcl} (c). 
 This indicates that the electron correlation effects 
 significantly promote the dimerization
 and lead the system to the insulating state. 
 Therefore, the insulating state realized for $U\ge U_c$ 
 is referred to as the dimer-insulator. 
 It is noted that the dimerization induced 
 by the electron correlation is 
  due to the effects of the off-diagonal 
    elements of the self-energy shift
  $\frac{\hat{\Sigma}(i\pi T)+\hat{\Sigma}^{\dagger}(i\pi T)}{2}$ which
  are the non-local effects neglected in the single-site DMFT and
 whose intra-dimer component is significantly enhanced in the present
 case as shown in Figs. \ref{fig_docc_2} (c) and (d). 

The enhancement of the dimerization is also observed in 
the two-particle quantities such as the
spin-spin correlation functions $\langle S_{mz}S_{m'z} \rangle$. 
The nearest neighbor spin-spin correlations  
 $\langle S_{mz}S_{m'z} \rangle$
 are shown in Figs. \ref{fig_tcl} (d) and (e). 
It is found that 
both the intra-plaquette component 
$\langle S_{1z}S_{2z} \rangle$  
 and the intra-dimer component 
$\langle S_{1z}S_{5z} \rangle$  
 are negative, that is, 
the nearest neighbor 
spin-spin correlations have tendency towards spin-singlet.
 With increasing $U$, the singlet correlation on a plaquette develops
  only slightly, while that on a dimer is largely enhanced. 
 We find that the characteristic features observed in $U$-dependence of
 the double occupancy $d$ are more clearly seen in that of 
 the intra-dimer  spin-spin correlation  
$\langle S_{1z}S_{5z} \rangle$.   
In particular, 
the sign-reversal of the curvature 
across the critical point $U_c$ 
in  $U$-$\langle S_{1z}S_{5z} \rangle$ curves 
 is more pronounced compared with $U$-$d$ curves  
 for $t_1>0.5t_2$. 
 
\begin{figure} [t]
  \begin{center}
   \includegraphics[width=5.5cm]{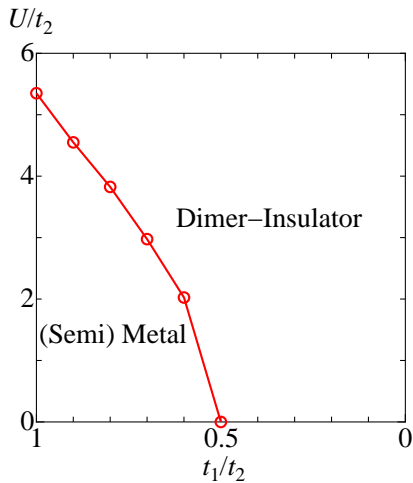} 
   \caption{(Color online) Phase diagram on $t_1$-$U$ plane for $T=0.1 t_2$. 
 \label{fig_pd}}
  \end{center}
 \end{figure}

Finally, we show the phase diagram 
on $t_1$-$U$ plane in Fig.  \ref{fig_pd}. For $t_1 \ge 0.5t_2$, 
the continuous transition between 
the (semi)metal and the dimer-insulator takes place at $U=U_c$ as
mentioned before, 
where $U_c$ is determined as the values at which
the Fermi surfaces vanish. The phase diagram is very similar to that
obtained previously by the 4-site CDMFT study\cite{Yanagi} and the
cluster-size dependence is small.   
 It should be noted that the critical value $U_c$ 
  is always
 smaller than the total bandwidth 
 $W=4t_1+2t_2$. 
 On the other hand, for $t_1<0.5t_2$, with increasing $U$, 
the electronic states smoothly vary from the noninteracting limit to
the strong coupling regime. 
 It is clear that the dimer-insulator is adiabatically connected to the
band insulator. 
 
\section{Summary and discussion}
 In summary, we have studied the nonmagnetic MIT in the Hubbard model 
on the 1/5-depleted square lattice at
half-filling for the case that the intra-dimer hopping $t_2$ is larger
than the intra-plaquette one $t_1$ by 
using the 8-site cellular dynamical mean field theory combined
with the continuous-time auxiliary field quantum Monte Carlo method.
 We have found that for $0.5t_2<t_1<t_2$, 
 the Coulomb interaction drives the MIT without any signatures of the
 first order transition, that is, the Mott transition is continuous 
  in contrast to the case of the ordinary square
 lattice Hubbard model in which the first order Mott
 transition 
 takes place. We have also shown that the continuous Mott transition 
 is nothing but a Lifshitz transition\cite{Lifshitz} and is originated from 
 the significant enhancement of the dimerization 
 by the non-local electron correlation effects\cite{Mazurenko,Poteryaev}. 
  These results are consistent with the CDMFT studies discussed for the mechanism of
 the MIT in NaV$_2$O$_5$\cite{Mazurenko} and
 Ti$_2$O$_3$\cite{Poteryaev}
 but the correlation-induced dimerization is more pronounced 
in the present case than the above two cases in which
 the inter-site interaction $V$ is essential for the formation of the
 sufficiently large band gap.
   Moreover, it is found that the dimer-insulator is adiabatically
 connected to the band insulator. 

  Finally, we comment on the stability of 
  the nonmagnetic continuous Mott
  transition against the antiferromagnetism. 
  In the present study,  
  our numerical calculations are performed within the
  paramagnetic phase and the possibility of the
  antiferromagnetism has not been considered so far.  
  Of course, the bipartite nature of the 1/5-depleted
  square lattice leads to the perfect nesting of the Fermi surfaces for
  $ t_1 \ge 0.5t_2$ as shown in Figs. \ref{fig_model} (c)-(f) and 
   the resulting bare staggered susceptibility diverges at $T=0$
  although the singularity is weaker than that 
  in the ordinary square lattice.  
  Therefore, whether the quantum fluctuations 
  remove this instability or not is the central issue.  
  Since in the 1/5-depleted square lattice, 
  the connectivity $z$ is small ($z=3$) and furthermore,  
  the lattice can be viewed as
   a collection of local units which naturally accommodate local
  singlets, such as 
  dimers or plaquettes, 
  the effects of quantum fluctuations are expected to be strong. 
  In fact, according to the results of the Heisenberg model 
  on the 1/5-depleted square lattice\cite{Ueda,Troyer,Troyer_2}, 
  the antiferromagnetism is
  confined only in the very small parameter region 
  $0.77t_2 \lesssim t_1 \lesssim 1.03t_2$\cite{Troyer,Troyer_2}.
  On the other hand, 
  the recent determinant QMC study 
  of the present model predicts 
  the AFI ground state for 
  the intermediate correlation regime\cite{khatami}. 
  Therefore, the nonmagnetic metallic states and 
  the continuous Mott transition observed in the present study 
  is considered to be the metastable state and 
  would be masked by the AFI state if one allows magnetic solutions. 
   However, because the quantum fluctuations in the present model 
  are  strong as mentioned above,  
  the nonmagnetic metallic states and 
  the continuous Mott transition can be stabilized
  by introducing a weak geometrical frustration. 
\begin{acknowledgments}
 A part of the
 computation in this work has been done using the facilities of the Supercomputer Center at ISSP, University
of Tokyo. This work has been supported by a Grant-in-Aid for Scientific Research on Innovative Areas Heavy
Electrons (No. 20102008) and (C) (No. 25400357).
\end{acknowledgments}



\begin{thebibliography}{99}
 \bibitem{Imada} M. Imada, A. Fujimori and Y. Tokura:
	 Rev. Mod. Phys. \textbf{70}, 1039 (1998).
 \bibitem{Meng} Z. Y. Meng, T. C. Lang, S. Wessel, F. F. Assaad and
	 A. Muramatsu,  
	 Nature (London) \textbf{464}, 847 (2010). 
\bibitem{Sorella} S. Sorella, Y. Otsuka and S. Yunoki, 
	 Sci. Rep. \textbf{2}, 992 (2012).
\bibitem{Kancharla} S. S. Kancharla and S. Okamoto,  
	Phys. Rev. B \textbf{75}, 193103 (2007). 
\bibitem{Ruger} R. R\"{u}ger, L. F. Tocchio, R. Valent\'{i} and C. Gros, 
	 	New J. Phys. \textbf{16} 033010 (2014).
\bibitem{Garg} A. Garg, H. R. Krishnamurthy and M. Randeria, 
	Phys. Rev. Lett. \textbf{97}, 046403 (2006).
\bibitem{Paris}N. Paris, K. Bouadim, F. Hebert, G. G. Batrouni and R. T. Scalettar,  
	Phys. Rev. Lett. \textbf{98}, 046403 (2007).
\bibitem{Kancharla_2} S. S. Kancharla and E. Dagotto, 
	Phys. Rev. Lett. \textbf{98}, 016402 (2007).
\bibitem{Ueda} K. Ueda, H. Kontani, M. Sigrist and P. A. Lee, 
	Phys. Rev. Lett. \textbf{76}, 1932 (1996).
\bibitem{Troyer} M. Troyer, H. Kontani, K. Ueda, 
	Phys. Rev. Lett. \textbf{76}, 3822 (1996).
\bibitem{Troyer_2} M. Troyer, M. Imada, K. Ueda, 
	J. Phys. Soc. Jpn. \textbf{66}, 2957 (1997).
\bibitem{Yamashita} Y. Yamashita, M. Tomura, Y. Yanagi and K. Ueda, 
 Phys. Rev. B \textbf{88}, 195104 (2013). 
\bibitem{Yanagi} Y. Yanagi and K. Ueda,  to be published in
	JPS Conf. Proc.  
\bibitem{Kotliar} G. Kotliar, S. Y. Savrasov, G. P\'{a}lsson and G. Biroli, 
Phys. Rev. Lett. \textbf{87}, 186401 (2001).
\bibitem{Maier} T. Maier, M. Jarrell, T. Pruschke and M. H. Hettler, 
Rev. Mod. Phys. \textbf{77}, 1027 (2005).
\bibitem{Gull_1} E. Gull, P. Werner, O. Parcollet and M. Troyer, 
Europhys. Lett. \textbf{82}, 57003 (2008).
\bibitem{Gull_2} E. Gull, A. J. Millis, A. I. Lichtenstein, A. N. Rubtsov, M. Troyer and P. Werner, 
Rev. Mod. Phys. \textbf{83}, 349 (2011).
 \bibitem{Park} H. Park, K. Haule and G. Kotliar,  
 	Phys. Rev. Lett. \textbf{101}, 186403  (2008).
\bibitem{Lifshitz} I. M. Lifshitz,  Sov. Phys. JETP \textbf{11}, 1130 (1960).
\bibitem{Mazurenko} V. V. Mazurenko, A. I. Lichtenstein, M. I. Katsnelson, 
	I. Dasgupta, T. Saha-Dasgupta and V. I. Anisimov, 
	Phys. Rev. B \textbf{66}, 081104 (2002).
\bibitem{Poteryaev} A. I. Poteryaev, A. I. Lichtenstein, and G. Kotliar, 
	Phys. Rev. Lett. \textbf{93}, 086401 (2004).
\bibitem{khatami} E. Khatami, R. R. P. Singh, W. E. Pickett, and 
	R. T. Scalettar, arXiv:1404.3731.
\end{thebibliography}
\end{document}